\documentclass[twocolumn]{autart}    

\usepackage{graphicx}          
\usepackage{cite}
\usepackage{bm}
\usepackage{amsmath,amssymb}
\usepackage{mathrsfs}
\usepackage{amsfonts}
\usepackage{color}

\begin{document}

\begin{frontmatter}

\title{Noise Suppression via Coherent Quantum Feedback: a Schr{\"o}dinger Picture Approach\thanksref{footnoteinfo}} 

\thanks[footnoteinfo]{This paper was not presented at any IFAC 
meeting. Corresponding author: Guofeng Zhang.}

\author[HK]{Shikun Zhang}\ead{zhang.shikun@outlook.com},    
\author[HK,SZ]{Guofeng Zhang}\ead{guofeng.zhang@polyu.edu.hk}

\address[HK]{Department of Applied Mathematics, The Hong Kong Polytechnic University, Hung Hom, Kowloon, Hong Kong Special Administrative Region, People's Republic of China}  
\address[SZ]{Shenzhen Research Institute, The Hong Kong Polytechnic University, Shenzhen 518057, China}


\begin{abstract}                          
In this article, we explore the possibility of achieving noise suppression for finite-dimensional quantum systems through coherent feedback. For a quantum plant which is expected to evolve according to a target trajectory, noise effect potentially deforms the plant state trajectory from the desired one. It is then hoped that a coherent feedback protocol can be designed that counteracts noise. In terms of coping with transient noise, we present several conditions on coherent feedback protocols under which noise-affected trajectories can be driven back towards desired ones asymptotically. As for rejecting persistent noise, conditions on protocols are given which ensure that the error between the target and feedback-corrected trajectories in the long-time limit can be effectively suppressed. Moreover, a possible construction of coherent feedback protocols which satisfies the given conditions is provided. Our theoretical results are illustrated by an example which involves a two-qubit plant and a two-level controller.
\end{abstract}

\end{frontmatter}

\section{Introduction}
It has long been realized that quantum systems may be harnessed to seek potential applications in, for instance, computing \cite{nielsen_chuang_2010,365700,10.1145/237814.237866,PhysRevLett.103.150502}, communication \cite{BENNETT20147,PhysRevLett.68.557,doi:10.1126/science.aan3211} and metrology \cite{toth2014quantum,RevModPhys.90.035005}. In this regard, quantum systems can potentially be very useful. However, at the same time, quantum systems can also be fragile, i.e., they are prone to noises that can be ubiquitous in real-world applications. If a quantum system is affected by noise, the very feature, such as entanglement, that makes it distinctive may be degraded, which may further disqualify its application in information processing tasks. Therefore, it is of great importance to seek certain measures that suppress noise effects on quantum systems.

Noise rejection is, in fact, an intensively studied subject in Systems and Control science and engineering. For classical systems, a go-to move would be designing and implementing feedback. In a nutshell, noises push system trajectories off the ones we desire. To remedy the situation, after acquiring sufficient information about the systems' real-time states, controllers proceed to process this information and ``feed" certain actions ``back" to the systems, guiding state trajectories towards favourable directions. However, the feedback mechanism is not to be naively applied to quantum systems, for measurement, which is the action of acquiring information, may bring non-negligible disturbances to quantum systems.

Fortunately, the coherent quantum feedback mechanism \cite{PhysRevA.49.4110,PhysRevA.62.022108}  circumvents the complications brought by quantum measurement. In terms of coherent feedback, a quantum system named ``controller" interacts coherently with the quantum system of interest called ``plant" (the interaction may be specifically engineered), and the two systems evolve together. Then, the goal is to obtain favorable reduced dynamics of the plant system by designing their joint dynamics. It is clear that no measurement on the plant system is involved. Compared with feedback in the classical scenario, \textit{the concept of ``loops" in coherent quantum feedback becomes vague.} In fact, any disturbance affecting the state of the quantum plant may affect the joint plant-controller state. The latter then evolves according to engineered joint dynamics, which in turn may regulate the plant's reduced dynamics. Intuitively, this mechanism can be viewed as ``controlling quantum systems via other quantum systems".

Since its introduction \cite{PhysRevA.49.4110,PhysRevA.62.022108}, coherent quantum feedback has received much attention (article \cite{DONG2022243} has included a review on the topic). Existing research related to coherent quantum feedback includes, to name a few, quantum LQG synthesis \cite{NURDIN20091837}, the SLH formalism \cite{5286277}, transfer function matrix realization \cite{PETERSEN20111757}, optimization of quantum dot conductance \cite{PhysRevB.90.205436}, linear fractional representation approach to quantum controller synthesis \cite{SICHANI2017264}, quantum coherent feedback network \cite{ZHANG2020108978} and robust control \cite{GZP19, DZA19a,GZP20,XIANG20178,9551752,dong2023h}.

In terms of existing research relevant with noise suppression with coherent feedback, article \cite{4625217} considers finding an $H^\infty$ quantum controller for linear quantum systems \cite{nurdin2017linear,ZHANG2022274}, which are quantum harmonic oscillators whose Heisenberg picture dynamics obey linear equations. The controller bounds the effect of the “energy” in the noise signal on the “energy” in the error signal. Also, article \cite{PhysRevA.78.032323} presents an experimental realization of the coherent feedback system proposed in \cite{4625217}. $H^\infty$ synthesis for linear quantum systems has also been covered in \cite{5648449}, with direct coupling considered. In \cite{PhysRevA.86.052304}, coherent feedback has been adopted to suppress decoherence for non-Markovian bosonic systems. It is shown that noise spectrum can be modulated by feedback, creating a noise-free band that includes the system characteristic frequency. Article \cite{PhysRevA.87.032117} shows that coherent feedback leads to a slower decay of coherence function of an atom in an optical cavity.

Moreover, through the introduction of a chaotic coherent feedback loop, article \cite{PhysRevA.92.033812} proposes a method that decouples the nanomechanical resonator from the environmental noise, in terms of optomechanical systems. In \cite{PhysRevA.97.062341}, coherent feedback has been applied in engineering system dynamics of a qubit that globally converges to target states tuned by control parameters, which can be viewed as a state preparation scheme against initial noise. In \cite{PhysRevA.99.053809}, it is shown that time-delayed coherent feedback is able to substantially preserve the coherence of a two-level system in a phononic cavity, even if finite temperature is considered. A coherent feedback protocol that involves sequential identical interactions has been introduced in \cite{PhysRevA.104.052614}, which results in discrete-time convergence to target states regardless of initial noise. Also, in \cite{LIU2022110236}, the designing of a a fault-tolerant $H^\infty$ controller for an optical parametric oscillator that admits time-varying uncertainties have been considered.

The above review is certainly not exhaustive. With noise suppression viewed, in a broader sense, as keeping unwanted effects on a system down, our message is that coherent feedback has been adopted to tackle a range of problems within this scope. However, to the best of our knowledge, the following problem has not been considered in other existing literature.Consider a finite-dimensional quantum plant which is expected to evolve following a designated pure state trajectory. If this plant is subject to noise, then is it possible to devise a coherent feedback protocol that counteracts the noise effect, in the sense that plant trajectories driven by both noise and feedback can be made close to the desired trajectory? Specifically, if the noise is transient, can we make the error between feedback-regulated and target trajectories asymptotically vanish? Moreover, in the case where the noise is persistent, can we push the error magnitude below a certain bar whose value can be actively tuned, given sufficient time for feedback action to weigh in?
 
With these questions in mind, we consider the following physical setup. Following our explanation on coherent feedback, the design of such protocols amounts to designing specific plant-controller composite systems. In this work, let us restrict to the case where composite systems admit Markovian evolution, which is described by (possibly time-varying) Lindblad master equations. The plant Hamiltonian has already been designated (which corresponds to the desired trajectory), and the noise coupling operators acting on the plant is assumed to be undesignable. Then, the parameters to be designed includes the (possibly time-varying) plant-controller interaction Hamiltonian and coupling operators acting on the controller. It is the primary goal of this work to explore the conditions on these parameters to be designed, under which satisfactory control performance can be achieved.  

To illustrate the relation of our work to existing literature, a comparison is in order. As long as coherent quantum feedback is considered for achieving certain goals, be it noise suppression or optimal control, our guiding principle and theirs should be similar: we are all coupling original systems with others and designing the resulting composite systems to fulfill certain objectives. However, to our belief, the problem studied in this article is a different one, and we are designing composite systems for different purposes. For example, the $H^\infty$ synthesis problem in \cite{4625217} concerns fulfilling an inequality involving \textit{observables} (and thus the Heisenberg picture is convenient) of infinite-dimensional quantum systems, and this article considers yielding certain \textit{state} trajectories (thus the Schr{\"o}dinger picture is preferred) of finite-dimensional systems. In this regard, our work may be supplementary to existing research on coherent feedback by expanding the utility of the mechanism.

Moreover, we note that there have also been works on noise reduction via measurement-based feedback \cite{PhysRevA.51.4913, PhysRevLett.79.2442,PhysRevA.60.1687}. The comparison between coherent and measurement-based feedback \cite{PhysRevX.4.041029,balouchi2017coherent} may be a complex issue itself, which is beyond the scope of this article. However, we note that, in \cite{PhysRevA.62.022108}, it is mentioned that in the ``conventional picture", sensors tend to destroy coherence and the controller is processing classical information. However, in the coherent feedback picture, the controller interacts coherently with the system and processes quantum information. We believe it remains to be seen whether measurement-based feedback can solve the task in this article and how its performance on this specific task compares to that of coherent feedback.


The remaining sections are organized as follows. In Section II, we state the plant-controller setup and describe the noise suppression problem. Section III covers the conditions on coherent feedback protocols regarding noise suppression, and we present an explicit protocol design which satisfies the given conditions. Section IV presents an example and Section V concludes the article.

\textit{Notations.} For any finite-dimensional Hilbert space $\mathcal{H}$, $\mathcal{B}(\mathcal{H})$ denotes the set of linear operators on $\mathcal{H}$, $\mathcal{B}_0(\mathcal{H})$ denotes the set of linear operators with zero trace on $\mathcal{H}$, and $\mathcal{D}(\mathcal{H})\subseteq \mathcal{B}(\mathcal{H})$ represents the set of positive semidefinite linear operators with trace one on $\mathcal{H}$. For any two finite-dimensional Hilbert spaces $\mathcal{H}_1$ and $\mathcal{H}_2$, let $\mathcal{H}_1 \otimes \mathcal{H}_2$ be their tensor product space, then $\text{tr}_{\mathcal{H}_j}(\cdot)$ denotes the partial trace operation over $\mathcal{H}_j$, $j=1,2$. When referring to operators in $\mathcal{B}(\mathcal{H})$, $\Vert \cdot \Vert$ represents trace norm, and for superoperators on $\mathcal{B}(\mathcal{H})$, $\Vert \cdot \Vert$ represents the norm induced by trace norm.

\section{Problem Description}
Let us consider a finite-dimensional quantum system on Hilbert space $\mathcal{H}_P$, which shall be referred to as the ``plant". Suppose that it is possible to engineer a plant Hamiltonian $H_P$, which is a Hermitian operator on $\mathcal{B}(\mathcal{H}_P)$. Ideally, we desire the evolution trajectory described by the following Liouville equation with designated initial value:
\begin{equation}\label{plant}
    \dot{\rho}=-\text{i}[H_P,\rho],\quad \rho (0)=|\phi_0\rangle\langle \phi_0|,
\end{equation}
where $|\phi_0\rangle \in \mathcal{H}_P$. Clearly, the solution to (\ref{plant}) is:
\begin{equation}\label{planttra}
    \rho_D(t)=e^{-\text{i}H_P t}|\phi_0\rangle\langle \phi_0| e^{\text{i}H_P t},\quad t \geq 0.
\end{equation}
That is, we want the pure state trajectory (\ref{planttra}) which is contained in $\mathcal{D}(\mathcal{H}_P)$. Theoretically, such a trajectory is straightforward to obtain: just simply initialize the plant in pure state $|\phi_0\rangle\langle \phi_0|$ and let it evolve with engineered Hamiltonian $H_P$. However, the situation turns out to be more complicated. 

Firstly, it should be noted that the plant may be subject to \textit{initialization error} and \textit{transient noise}. Here, initialization error means that the initialized plant state is not perfectly equal to $|\phi_0\rangle\langle \phi_0|$. From a physics perspective, any environmental interaction with the plant can be viewed as transient noise if it possibly pushes the plant state off the desired trajectory and persists only for a finite amount of time. Mathematically speaking, if only these two scenarios are considered, then there exists $t_a \geq 0$ and $\rho_a \in \mathcal{D}(\mathcal{H}_P)$, such that 
\begin{equation}\label{planttra1}
    \rho(t)=e^{-\text{i}H_P t}\rho_a e^{\text{i}H_P t},\quad t \geq t_a.
\end{equation}
It is clear that trajectories (\ref{planttra}) and (\ref{planttra1}) may not be the same. Also, initialization error may be viewed as transient noise which acts in the beginning of the evolution. We would like to note that transient noise described here is a standard consideration in quantum information literature. The foundational book \cite{nielsen_chuang_2010} has covered noises that align with our description.

Secondly, the plant may be affected by \textit{persistent noise}. In this case, the noise may persistently act on the plant, potentially invalidating the plant model (\ref{plant}). A possible new model for this case is described by the following Lindblad master equation:
\begin{multline}\label{plant_persistantn}
    \dot{\rho}=-\text{i}[H_P,\rho]\\
    +\sum_{k=1}^M L_{P,k} \rho L_{P,k}^\dagger -\frac{1}{2}L_{P,k}^\dagger L_{P,k} \rho -\frac{1}{2}\rho L_{P,k}^\dagger L_{P,k},
\end{multline}
where $L_k \in \mathcal{B}(\mathcal{H}_P)$ is called a \textit{coupling operator} ($1\leq k \leq M$), which describes the environmental noise effect on the plant. In other words, with persistent noise acting on the plant, we might not even be able to use our original model. Note that model (\ref{plant_persistantn}) effectively describes the dynamics of quantum systems weakly coupled to Markovian reservoirs, for instance, qubits in bosonic fields. It involves the physical assumption that environmental excitations caused by the system fades quickly. We refer to the book \cite{10.1093/acprof:oso/9780199213900.001.0001} regarding such models.

To remedy the situation, we propose to take the approach of ``coherent quantum feedback". Let us consider another finite-dimensional quantum system on Hilbert space $\mathcal{H}_C$, which is referred to as the ``controller". If the controller interacts with the plant (while the plant-controller composite system evolves as a whole), an action on the plant is induced. What we hope is that, with specifically designed controllers and plant-controller interactions, such actions may suppress the effect of noise on the plant. 

We then mathematically formulate the idea. The plant-controller composite system is built upon the Hilbert space $\mathcal{H}_P \otimes \mathcal{H}_C$. We seek to design a possibly time-dependent plant-controller interaction Hamiltonian $H_I (t)$, which is viewed as a function from $[0,+\infty)$ to the Hermitian subset of $\mathcal{B}(\mathcal{H}_P \otimes \mathcal{H}_C)$, and a set of couplings operators $\{I_P \otimes L_{C,k}\}_{k=1}^N \subseteq \mathcal{B}(\mathcal{H}_P \otimes \mathcal{H}_C)$ that act nontrivially on the controller only ($I_P$ denotes the identity operator on $\mathcal{H}_P$), with which the composite plant-controller system dynamics is governed by the following (possibly) time-dependent Lindblad master equation:
\begin{equation}\label{plant_controller1}
    \dot{\sigma}=\big(\mathcal{L}_{\text{p}} + \mathcal{L}_{\text{fb}}(t)+\mathcal{L}_{\text{noise}}\big)(\sigma),
\end{equation}
with superoperators $\mathcal{L}_{\text{p}}$, $\mathcal{L}_{\text{fb}}(t)$ and $\mathcal{L}_{\text{noise}}$ defined as:
\begin{equation}\label{plant_controller2}
    \mathcal{L}_{\text{p}}(\cdot)\triangleq -\text{i}[H_P\otimes I_C, \cdot],
\end{equation}
\begin{multline}\label{plant_controller3}
    \mathcal{L}_{\text{fb}}(t)(\cdot)\triangleq -\text{i}[H_I (t), \cdot]+\sum_{k=1}^N(I_P \otimes L_{C,k})(\cdot)(I_P \otimes L_{C,k}^\dagger)\\
    -\frac{1}{2}(I_P \otimes L_{C,k}^\dagger L_{C,k})(\cdot)-\frac{1}{2}(\cdot)(I_P \otimes L_{C,k}^\dagger L_{C,k}),
\end{multline}
\begin{multline}\label{plant_controller4}
    \mathcal{L}_{\text{noise}}(\cdot)\triangleq \sum_{k=1}^M(L_{P,k}\otimes I_C)(\cdot)(L_{P,k}^\dagger \otimes I_C)\\
    -\frac{1}{2}(L_{P,k}^\dagger L_{P,k} \otimes I_C)(\cdot)-\frac{1}{2}(\cdot)(L_{P,k}^\dagger L_{P,k} \otimes I_C),
\end{multline}
where $I_C$ stands for the identity operator on $\mathcal{H}_C$ \big(also, recall $L_{P,k}$ from (\ref{plant_persistantn}) \big). 

Clearly, $\mathcal{L}_{\text{noise}}$ represents persistent noise. The effect of transient noise is modeled as follows. Suppose that at time $t_a$ the system state is pushed to some $\rho'_a  \in \mathcal{D}(\mathcal{H}_P \otimes \mathcal{H}_C)$ by transient noise. Then, the noise-affected state trajectory after $t_a$ is taken to be the solution of (\ref{plant_controller1}) which passes through $(t_a,\rho'_a)$ on the time interval $(t_a,+\infty)$. 

Next, we note that coherent feedback is expected to suppress external noise, not to be a noise source itself. This means that if the initial plant state is perfectly set in $|\phi_0\rangle\langle \phi_0|$ \big(eq.(\ref{plant})\big) and external noise is absent, the implementation of coherent feedback should not alter the plant's desired evolution. This requirement should be considered in designing coherent feedback.

Consider any solution, say $\sigma (t)$, of (\ref{plant_controller1}) which passes through some $(t_a, \rho''_a)$, with $t_a \geq 0$ and $\rho''_a \in \mathcal{D}(\mathcal{H}_P \otimes \mathcal{H}_C)$. It is clear that $\text{tr}_{\mathcal{H}_C}\big(\sigma (t)\big)$ ($t\geq t_e$) represents the plant's state trajectory which is possibly affected by noise but, at the same time, possibly remedied by coherent feedback protocol. We expect that the coherent feedback fulfills the following requirements:

(i) In the absence of noise and with the plant initially residing in state $|\phi_0\rangle\langle \phi_0|$, the evolution of the plant state follows the desired trajectory while coherent feedback is being implemented. That is, $\text{tr}_{\mathcal{H}_C}\big(\sigma (t)\big)=\rho_D (t)$, $t \geq 0$, where $\rho_D (t)$ is given in (\ref{planttra}).

(ii) In the case where only initialization error and transient noise are considered ($\mathcal{L}_{\text{noise}}=0$), the error between plant state trajectory and desired trajectory (\ref{planttra}) asymptotically vanishes, i.e., $\lim_{t\to +\infty} \text{tr}_{\mathcal{H}_C}\big(\sigma (t)\big)-\rho_D (t)=0$. 

(iii) If persistent noise is present ($\mathcal{L}_{\text{noise}}\neq 0$), then the upper bound of $\limsup_{t\to +\infty} \Vert \text{tr}_{\mathcal{H}_C}\big(\sigma (t)\big)-\rho_D (t)\Vert$ can be made arbitrarily small by designing suitable $\mathcal{L}_{\text{fb}}(\cdot)$. This indicates that the error magnitude approaches $[0,a]$, where $a>0$ can be made infinitely close to 0 by coherent feedback design.  

\section{Main Results}
As main results of this article, we present a few conditions on coherent feedback design in this section, under which the desired performance mentioned in the previous section can be achieved. Also presented is a possible construction that satisfies the given conditions.

\subsection{Fulfilling Requirement (i)}

In this subsection, we present the conditions under which the first requirement can be satisfied. The result is given by the following theorem.

\begin{thm}
Suppose that $\mathcal{L}_{\emph{noise}}=0$. If there exists $\rho_C \in \mathcal{D}( \mathcal{H}_C)$, such that $\mathcal{L}_{\emph{fb}}(0)\big(|\phi_0\rangle\langle\phi_0| \otimes \rho_C\big)=0$, and if
\begin{equation}\label{TDH}
    H_I (t)=e^{-\emph{i}(H_P\otimes I_C)t}H_I (0) e^{\emph{i}(H_P\otimes I_C)t}, \quad t \geq 0,
\end{equation}
then with $\sigma (0)=|\phi_0\rangle\langle\phi_0| \otimes \rho_C$, it holds that
\begin{equation}\label{thm1}
    \emph{tr}_{\mathcal{H}_C}(\sigma (t))=\rho_D (t),\quad t \geq 0.
\end{equation}
\end{thm}

\begin{pf}
Let $\sigma (\cdot)$ be a solution of (\ref{plant_controller1}). Denote
\begin{equation}\label{Up}
    U_P(t)\triangleq e^{-\text{i}(H_P\otimes I_C)t},\quad t \in \mathbb{R},
\end{equation}
and
\begin{equation}\label{theta}
    \theta (t)\triangleq U_P^\dagger (t) \sigma (t) U_P (t), \quad t \in \mathbb{R}.
\end{equation}
We shall prove that 
\begin{equation}\label{theta1}
    \dot{\theta}=\mathcal{L}_{\text{fb}}(0)\big(\theta \big).
\end{equation}
After that, we will show that, if the unitarily transformed trajectory (\ref{theta}) fulfills (\ref{theta1}), then $\sigma (\cdot)$ fulfills (\ref{thm1}).

It is clear that 
\begin{equation}\label{t1}
    \dot{\theta}=\dot{U}_P^\dagger \sigma U_P +U_P^\dagger \dot{\sigma} U_P +U_P^\dagger \sigma \dot{U}_P.
\end{equation}
The terms on the r.h.s of (\ref{t1}) will be computed separately, and the results will be combined to prove (\ref{theta1}).

Since $U^{\dagger}_P$ and $H_P\otimes I_C$ commute, we compute that
\begin{equation}\label{t2}
\begin{aligned}
\dot{U}_P^\dagger \sigma U_P&=\text{i}U_P^\dagger (H_P\otimes I_C) \sigma U_P\\
                            &=\text{i}(H_P\otimes I_C) U_P^\dagger  \sigma U_P\\ 
                            &=\text{i}(H_P\otimes I_C) \theta.\\
\end{aligned}
\end{equation}
Conjugating both sides of (\ref{t2}) yields:
\begin{equation}\label{t3}
U_P^\dagger \sigma \dot{U}_P=-\text{i}\theta (H_P \otimes I_C) 
\end{equation}
With (\ref{t2}) and (\ref{t3}), we have
\begin{equation}\label{t4}
    \dot{U}_P^\dagger \sigma U_P+U_P^\dagger \sigma \dot{U}_P=\text{i}[H_P\otimes I_C,\theta].
\end{equation}
To evaluate $U_P^\dagger \dot{\sigma} U_P$, eqs. (\ref{plant_controller1})-(\ref{plant_controller3}) are referred to. The terms there are dealt with separately.

Considering that $U_P$ and $H_P\otimes I_C$ commute, it can be verified that, 
\begin{equation}\label{t5}
    \begin{aligned}
        &-\text{i}U_P^\dagger  [H_P \otimes I_C,\sigma]U_P\\
        =&-\text{i}\big(U_P^\dagger H_P \otimes I_C \sigma U_P -U_P^\dagger \sigma H_P \otimes I_C U_P\big)\\
        =&-\text{i} [H_P \otimes I_C, \theta].
    \end{aligned}
\end{equation}
With (\ref{TDH}), for $t \geq 0$,
\begin{equation}\label{t6}
    \begin{aligned}
        &-\text{i}U_P^\dagger (t) [H_I (t),\sigma (t)]U_P (t)\\
        =&-\text{i}\big(U_P^\dagger (t) H_I (t)\sigma (t) U_P (t) -U_P^\dagger (t) \sigma (t) H_I (t) U_P (t)\big)\\
        =&-\text{i}\big(H_I (0) \theta(t) -\theta(t) H_I (0) \big).
    \end{aligned}
\end{equation}
Also, since $U_P$ acts trivially on $\mathcal{H}_C$, we have
\begin{equation}\label{t7}
    \begin{aligned}
        &U_P^\dagger \sum_{k=1}^N(I_P \otimes L_{C,k})\sigma(I_P \otimes L_{C,k}^\dagger) U_P\\
        =&\sum_{k=1}^N(I_P \otimes L_{C,k})\theta (I_P \otimes L_{C,k}^\dagger),
    \end{aligned}
\end{equation}

\begin{equation}\label{t8}
        U_P^\dagger \sum_{k=1}^N(I_P \otimes L_{C,k}^\dagger L_{C,k})\sigma U_P=\sum_{k=1}^N(I_P \otimes L_{C,k}^\dagger L_{C,k})\theta,
\end{equation}
and
\begin{equation}\label{t9}
        U_P^\dagger \sigma \sum_{k=1}^N(I_P \otimes L_{C,k}^\dagger L_{C,k})U_P=\sum_{k=1}^N\theta(I_P \otimes L_{C,k}^\dagger L_{C,k})
\end{equation}
\end{pf}
Combining (\ref{plant_controller1}), (\ref{t1}), (\ref{t4})--(\ref{t9}), we arrive at (\ref{theta1}).

Next, if $\sigma (0)=|\phi_0\rangle\langle\phi_0| \otimes \rho_C$, then $\theta (0)=|\phi_0\rangle\langle\phi_0| \otimes \rho_C$. Since $\mathcal{L}_{\text{fb}}(0)\big(|\phi_0\rangle\langle\phi_0| \otimes \rho_C\big)=0$, as given in the statement of the theorem, it is true that $\theta (t)=|\phi_0\rangle\langle\phi_0| \otimes \rho_C$, $t \geq 0$. As a result, 
\begin{equation}
    \sigma (t)=U_P (t)\theta (t) U_P^\dagger (t)=\rho_D (t)\otimes \rho_C,\quad t \geq 0.
\end{equation}
and thus
\begin{equation}
    \text{tr}_{\mathcal{H}_C}(\sigma (t))=\rho_D (t),\quad t \geq 0,
\end{equation}
which is (\ref{thm1}). The proof is completed. $\hfill\square$

{\bfseries Remark 1 }\textit{Theorem 1 provides certain conditions under which a coherent feedback protocol does not disrupt the desired trajectory (\ref{planttra}), given that no noise is present and the plant is perfectly initialized. That is: (i) the controller is initialized in} $\rho_C ;$\textit{ (ii)} $\mathcal{L}_{\text{fb}}(0)\big(|\phi_0\rangle\langle\phi_0| \otimes \rho_C\big)=0$ \textit{; (iii) the plant-controller interaction Hamiltonian follows (\ref{TDH}). The idea is that $ \text{tr}_{\mathcal{H}_C}(\sigma (t))=\rho_D (t)$ holds if the unitarily transformed trajectory (\ref{theta}) remains steady at $|\phi_0\rangle\langle \phi_0| \otimes  \rho_C$. Moreover, the realization of (\ref{TDH}) seems challenging at first glance. However, we present a discussion on its realization in Appendix B, which may open up a topic for future research.}

At this point, one may ask the question: is it possible that imperfect initialization of the controller state $\rho_C$ will affect the desired trajectory? Our answer is a frank yes. Such a possibility does exist. Readers are assured that this issue will be addressed in the remaining part of this article, but before doing that, we would like to present the following way of viewing this issue. In theory, a control protocol may be designed with the property of achieving ideal control performance, but in practice, it is normal that the implemented protocol may not exactly agree with the theoretically designed one. In the case of the latter, it would be unrealistic to expect that the control performance will not change at all. It suffices to guarantee that the real performance lies within the scope of tolerance, which is another possible property of the protocol to be designed.

As for this work, it is worthwhile mentioning that the controller's initial state is \textbf{a part of the coherent feedback protocol}. Therefore, imperfect controller initialization is only a part of all possible imperfections the protocol may have. By implying that a designed protocol has desired performance, Theorem 1 has fulfilled its job. The imperfections, including controller initialization error, will be addressed through Theorems 2 and 3. 

\subsection{Fulfilling Requirement (ii)}

We have not yet covered the issue of plant initialization error and transient noise. The following result is presented which resolves a part of the mentioned concern. 

\begin{thm}
Suppose that $\mathcal{L}_{\emph{noise}}=0$, and there exists $\rho_C \in \mathcal{D}( \mathcal{H}_C)$, such that $|\phi_0\rangle\langle\phi_0| \otimes \rho_C$ is the unique steady state of the system $\dot{y}=\mathcal{L}_{\emph{fb}}(0)\big(y\big)$ in $\mathcal{D}(\mathcal{H}_P \otimes \mathcal{H}_C)$, and also (\ref{TDH}) holds.
Then, $\forall t_0 \geq 0$ and $\forall \sigma_0 \in \mathcal{D}(\mathcal{H}_P \otimes \mathcal{H}_C)$, the solution of (\ref{plant_controller1}) which passes through $(t_0,\sigma_0)$, denoted by $\sigma(t;t_0,\sigma_0)$, satisfies that
\begin{equation}\label{thm2}
    \lim_{t\to +\infty }\emph{tr}_{\mathcal{H}_C}\big(\sigma(t;t_0,\sigma_0)\big)-\rho_D (t)=0.
\end{equation}
\end{thm}

\begin{pf}
Let us denote:
\begin{equation}\label{mu}
    \mu (t)\triangleq U_P^\dagger (t) \sigma(t;t_0,\sigma_0) U_P (t), \quad t \in \mathbb{R},
\end{equation}
with $U_P (t)$ given in (\ref{Up}). Then, according to the proof of Theorem 1, we have
\begin{equation}
    \dot{\mu}=\mathcal{L}_{\text{fb}}(0)\big(\mu \big).
\end{equation}
Again, we will harness the property of the unitarily transformed trajectory $\mu$ to prove the theorem.

Because $\sigma(t;t_0,\sigma_0)$ passes through $(t_0,\sigma_0)$,  $\mu(t)$ passes through $(t_0,U_P^\dagger (t_0)\sigma_0 U_P (t_0))$. If $\forall \sigma_0$, $\mu(t)$ converges to $|\phi_0\rangle\langle\phi_0| \otimes \rho_C$, which is the unique steady state of the system $\dot{y}=\mathcal{L}_{\text{fb}}(0)\big(y\big)$ in $\mathcal{D}(\mathcal{H}_P \otimes \mathcal{H}_C)$, then this theorem can be proved. Let us denote $U_P^\dagger (t_0)\sigma_0 U_P (t_0)$ by $\mu_{t_0}$, which is $\mu(t_0)$ according to (\ref{mu}). Then, $\forall t_0 \geq 0$ and $\forall t \geq t_0$, it holds that
\begin{multline}
    \mu(t)=e^{\mathcal{L}_{\text{fb}}(0)\cdot t}\mu (0)=e^{\mathcal{L}_{\text{fb}}(0)\cdot (t-t_0)} e^{\mathcal{L}_{\text{fb}}(0)\cdot t_0}\mu (0)\\
    =e^{\mathcal{L}_{\text{fb}}(0)\cdot (t-t_0)}\mu_{t_0}.
\end{multline}
Since $\sigma_0 \in \mathcal{D}(\mathcal{H}_P \otimes \mathcal{H}_C)$, by (\ref{mu}), $\mu_{t_0} \in \mathcal{D}(\mathcal{H}_P \otimes \mathcal{H}_C)$. Moreover, note that $|\phi_0\rangle\langle\phi_0| \otimes \rho_C$ is the unique steady state of $\mathcal{L}_{\text{fb}}(0)$ in $\mathcal{D}(\mathcal{H}_P \otimes \mathcal{H}_C)$, and $\mathcal{L}_{\text{fb}}(0)$ is the generator of a time-independent Lindblad master equation. We thus have 
\begin{equation}
    \lim_{t \to +\infty} e^{\mathcal{L}_{\text{fb}}(0)\cdot t}\mu_{t_0}=|\phi_0\rangle\langle\phi_0| \otimes\rho_C,
\end{equation}
according to \cite{PhysRevA.81.062306}. Therefore,
\begin{multline}\label{mulimit}
    \lim_{t \to +\infty}\mu(t)=\lim_{t \to +\infty}e^{\mathcal{L}_{\text{fb}}(0)\cdot (t-t_0)}\mu_{t_0}\\
    =\lim_{t \to +\infty} e^{\mathcal{L}_{\text{fb}}(0)\cdot t}\mu_{t_0}=|\phi_0\rangle\langle\phi_0| \otimes \rho_C.
\end{multline}
Next, it can be checked that
\begin{equation}\label{sigmalimitcycle}
    \begin{aligned}
        &\Vert \mu (t)-|\phi_0\rangle\langle\phi_0| \otimes \rho_C\Vert \\
        =&\Vert U_P (t) \big(\mu (t)-|\phi_0\rangle\langle\phi_0| \otimes \rho_C\big) U_P^\dagger (t)\Vert \\
        =&\Vert \sigma(t;t_0,\sigma_0)-\rho_D (t) \otimes \rho_C\Vert .
    \end{aligned}
\end{equation}
By (\ref{mulimit}), it is clear that $\lim_{t \to +\infty} \Vert \mu (t)-|\phi_0\rangle\langle\phi_0| \otimes \rho_C\Vert=0$. Therefore, according to (\ref{sigmalimitcycle}), we arrive at
\begin{equation}
   \lim_{t \to +\infty} \sigma(t;t_0,\sigma_0)-\rho_D (t) \otimes \rho_C=0,
\end{equation}
which indicates that 
\begin{equation}
    \lim_{t\to +\infty }\text{tr}_{\mathcal{H}_C}\big(\sigma(t;t_0,\sigma_0)\big)-\rho_D (t)=0
\end{equation}
$\hfill\square$
\end{pf}

{\bfseries Remark 2} \textit{Theorem 2 says that, if conditions mentioned in (ii) and (iii) in Remark 1 are satisfied, and in addition,} $|\phi_0\rangle\langle\phi_0| \otimes \rho_C$ \textit{is the unique steady-state density operator of} $\mathcal{L}_{\text{fb}}(0)$, \textit{then for \textbf{any} time $t_0$ greater or equal to 0, and \textbf{regardless} of what quantum state the plant-controller system resides in at time $t_0$, as long as the system continues to evolve according to (\ref{plant_controller1}) without persistent noise, the plant's state trajectory asymptotically approaches the desired one.}

\textit{By Theorem 2, no matter what kind of initialization error, or what kind of transient noise (e.g., plant, controller, or correlated noise) has occurred (and whenever it occurs), the distance between the plant's state trajectory and the desired one tends to zero, resolving the initialization error and transient noise issue and showing suppression of noise effect.}

\subsection{Fulfilling Requirement (iii)}
Let us now turn to the case where $\mathcal{L}_{\text{noise}}\neq 0$, i.e., noise persistently acts on the plant, so that the plant dynamics without coherent feedback is described by a Lindblad master equation. In this case, coherent feedback aims at continuously suppressing the noise effect.

Unlike the case where $\mathcal{L}_{\text{noise}}=0$, it is no longer expected that the plant state trajectory regulated by coherent feedback will still asymptotically approach trajectory (\ref{planttra}). Instead, we now hope that $\text{tr}_{\mathcal{H}_C}(\sigma (t))$ will asymptotically approach a tube-shaped region centered around $\rho_D (t)$, whose radius is controlled by the coherent feedback protocol. It is also hoped that there is a sequence of protocols under which the resulting radius tends to 0. If these goals can be achieved, then for any $\epsilon >0$, there is a time $T_{\epsilon}$ and a feedback protocol, under which 
\begin{equation}\label{hope}
    \Vert \text{tr}_{\mathcal{H}_C}\big(\sigma (t)\big)-\rho_D (t)\Vert <\epsilon, \quad t>T_{\epsilon}.
\end{equation}
In other words, we expect that the noise-affected plant trajectory can be corrected with arbitrary precision after sufficient amount of time. The following result shows that, with the same conditions in Theorem 2 except that now $\mathcal{L}_{\text{noise}}\neq 0$, it is possible to achieve (\ref{hope}).

\begin{thm}
Suppose that there exists $\rho_C \in \mathcal{D}( \mathcal{H}_C)$, such that $|\phi_0\rangle\langle\phi_0| \otimes \rho_C$ is the unique steady state of the system $\dot{y}=\mathcal{L}_{\emph{fb}}(0)\big(y\big)$ in $\mathcal{D}(\mathcal{H}_P \otimes \mathcal{H}_C)$, and also (\ref{TDH}) holds. Then, there exist $\alpha >0$ and $K>0$, such that when coherent feedback protocol $\gamma \mathcal{L}_{\emph{fb}}(\cdot)$ ($\gamma >0$) is implemented, it holds that
\begin{equation}\label{errorbound}
    \limsup_{t\to +\infty}\Vert \emph{tr}_{\mathcal{H}_C}\big(\sigma (t)\big)-\rho_D (t)\Vert \leq \frac{K}{\gamma \alpha} \Vert \mathcal{L}_{\emph{noise}}\Vert,
\end{equation}
for any $\sigma (0)\in \mathcal{D}(\mathcal{H}_P \otimes \mathcal{H}_C)$.
\end{thm}
\begin{pf}
Let us consider the trajectory
\begin{equation}
\lambda (t)\triangleq \rho_D (t)\otimes \rho_C.
\end{equation}
We shall prove that 
\begin{equation}\label{thm3_6}
    \dot{\lambda}=\big(\mathcal{L}_{\text{p}} + \gamma \mathcal{L}_{\text{fb}}(t)\big)(\lambda),
\end{equation}
for $\gamma >0$.
Note that $\lambda (t)$ is associated with the desired trajectory $\rho_D (t)$. With the differential equation of $\sigma$ already known \big(eq.(\ref{plant_controller1})\big), eq. (\ref{thm3_6}) will be applied in the derivation of the differential equation of $\sigma-\lambda$, which is crucial for bounding the error in (\ref{errorbound}).

Clearly, $\dot{\lambda}=\mathcal{L}_{\text{p}}(\lambda)$ holds. We shall prove that $\mathcal{L}_{\text{fb}}(t)\big)(\lambda)=0$ by dealing with each term of $\mathcal{L}_{\text{fb}}(t)$. It is checked that the following results hold. Firstly,
\begin{equation}\label{thm3_1}
    \begin{aligned}
        &-\text{i}[H_I (t),\lambda(t)]\\
        =&-\text{i}[U_P (t) H_I (0)U_P^\dagger (t),U_P (t) (|\phi_0\rangle\langle\phi_0| \otimes \rho_C) U_P^\dagger (t)]\\
        =&-\text{i}U_P (t) [H_I (0),|\phi_0\rangle\langle\phi_0| \otimes \rho_C]U_P^\dagger (t).
    \end{aligned}
\end{equation}
Secondly,
\begin{equation}\label{thm3_2}
    \begin{aligned}
        &\sum_{k=1}^N(I_P \!\otimes \! L_{C,k})\lambda(t)(I_P \! \otimes \!L_{C,k}^\dagger)\\
        =&\sum_{k=1}^N(I_P \! \otimes \! L_{C,k})U_P (t) (|\phi_0\rangle\langle\phi_0| \! \otimes \!\rho_C) U_P^\dagger (t)(I_P \! \otimes \! L_{C,k}^\dagger)\\
        =&U_P (t)\! \sum_{k=1}^N(I_P \!\otimes \! L_{C,k})(|\phi_0\rangle\langle\phi_0| \!\otimes \! \rho_C) (I_P \!\otimes \! L_{C,k}^\dagger)U_P^\dagger (t).
    \end{aligned}
\end{equation}
Thirdly,
\begin{equation}\label{thm3_3}
    \begin{aligned}
        &\sum_{k=1}^N(I_P \!\otimes \! L_{C,k}^\dagger L_{C,k})\lambda(t)\\
        =&\sum_{k=1}^N(I_P \!\otimes \! L_{C,k}^\dagger L_{C,k})U_P (t) (|\phi_0\rangle\langle\phi_0| \!\otimes \! \rho_C) U_P^\dagger (t)\\
        =&U_P (t)\!\sum_{k=1}^N(I_P \!\otimes \! L_{C,k}^\dagger L_{C,k}) (|\phi_0\rangle\langle\phi_0| \!\otimes \! \rho_C) U_P^\dagger (t).
    \end{aligned}
\end{equation}
Fourthly,
\begin{equation}\label{thm3_4}
    \begin{aligned}
        &\sum_{k=1}^N\lambda(t)(I_P \!\otimes \! L_{C,k}^\dagger L_{C,k})\\
        =&\sum_{k=1}^N U_P (t) (|\phi_0\rangle\langle\phi_0| \!\otimes \! \rho_C) U_P^\dagger (t)(I_P \!\otimes \! L_{C,k}^\dagger L_{C,k})\\
        =&U_P (t)\!\sum_{k=1}^N  (|\phi_0\rangle\langle\phi_0| \!\otimes \! \rho_C) (I_P \!\otimes \! L_{C,k}^\dagger L_{C,k})U_P^\dagger (t).
    \end{aligned}
\end{equation}
Combining (\ref{thm3_1}), (\ref{thm3_2}), (\ref{thm3_3}) and (\ref{thm3_4}), we arrive at
\begin{equation}\label{thm3_5}
\begin{aligned}
        &\mathcal{L}_{\text{fb}}(t)\big(\lambda(t)\big)=U_P (t) \mathcal{L}_{\text{fb}}(0)\big(|\phi_0\rangle\langle\phi_0| \otimes \rho_C\big)U_P^\dagger (t)\\
        &=0,
    \end{aligned}
\end{equation}
since $|\phi_0\rangle\langle\phi_0| \otimes \rho_C$ is a steady state of $\mathcal{L}_{\text{fb}}(0)$. Therefore, the equation (\ref{thm3_6}) holds for $\gamma >0$. If the coherent feedback protocol $\gamma \mathcal{L}_{\text{fb}}(\cdot)$ with $\gamma >0$ is implemented, then (\ref{plant_controller1}) is modified as:
\begin{equation}\label{plant_controller11}
    \dot{\sigma}=\big(\mathcal{L}_{\text{p}} + \gamma \mathcal{L}_{\text{fb}}(t)+\mathcal{L}_{\text{noise}}\big)(\sigma).
\end{equation}
By subtracting (\ref{thm3_6}) from (\ref{plant_controller11}) and denoting $E\triangleq \sigma-\lambda$, we have
\begin{equation}\label{thm3_7}
    \dot{E}=\big(\mathcal{L}_{\text{p}} + \gamma\mathcal{L}_{\text{fb}}(t)\big)(E)+\mathcal{L}_{\text{noise}}(\sigma).
\end{equation}
As a result, it holds that
\begin{equation}\label{thm3_8}
    E(t)=\mathcal{G}_{\gamma}(t,0)E(0)+\int_0^t \mathcal{G}_{\gamma}(t,s) \mathcal{L}_{\text{noise}}\big(\sigma(s)\big)ds,
\end{equation}
where $\mathcal{G}_{\gamma}(\cdot,\cdot)$ denotes the state transition superoperator of the following dynamical system:
\begin{equation}\label{thm3_9}
    \dot{z}=\big(\mathcal{L}_{\text{p}} + \gamma\mathcal{L}_{\text{fb}}(t)\big)(z).
\end{equation}
With (\ref{thm3_8}), we will first show that $\mathcal{G}_{\gamma}(t,0)E(0)$ tends to zero as $t$ tends to infinity.

Let us make the following notation:
\begin{equation}\label{thm3_10}
    U_P(t)(\cdot)U_P^\dagger (t)\triangleq \mathcal{U}_t (\cdot).
\end{equation}
We shall then prove that
\begin{equation}\label{thm3_12}
    \mathcal{G}_{\gamma}(t,s)=\mathcal{U}_t e^{\gamma\mathcal{L}_{\text{fb}}(0)\cdot (t-s)}\mathcal{U}_s^\dagger.
\end{equation}
For any $s\geq 0$ and any $a\in \mathcal{B}(\mathcal{H}_P\otimes\mathcal{H}_C)$, we denote the solution of (\ref{thm3_9}) that passes through $(s,a)$ as $A(\cdot;s,a)$. Also, we define $B(t)\triangleq \mathcal{U}_t^\dagger \big(A(\cdot;s,a)\big)$. Following the same proof that results in (\ref{theta1}), we also have $\dot{B}=\gamma\mathcal{L}_{\text{fb}}(0)\big(B\big)$. Then, for $t\geq s$, it is verified that
\begin{equation}\label{thm3_11}
    \begin{aligned}
        A(t;s,a)&=\mathcal{G}_{\gamma}(t,s)\big(a\big)\\
                &=\mathcal{U}_t \big(B(t)\big)\\
                &=\mathcal{U}_t e^{\gamma\mathcal{L}_{\text{fb}}(0)\cdot (t-s)}\big(B(s)\big)\\
                &=\mathcal{U}_t e^{\gamma\mathcal{L}_{\text{fb}}(0)\cdot (t-s)}\mathcal{U}_s^\dagger \big(A(s;s,a)\big)\\
                &=\mathcal{U}_t e^{\gamma\mathcal{L}_{\text{fb}}(0)\cdot (t-s)}\mathcal{U}_s^\dagger \big(a\big).
    \end{aligned}
\end{equation}
Therefore, due to the arbitrariness of $a$, (\ref{thm3_12}) holds.
Following (\ref{thm3_12}), we derive that
\begin{equation}\label{thm3_13}
    \mathcal{G}_{\gamma}(t,0)E(0)=\mathcal{U}_t e^{\gamma\mathcal{L}_{\text{fb}}(0)\cdot t}E(0).
\end{equation}
Since $\lambda (0)$ is the unique steady state of the system $\dot{y}=\mathcal{L}_{\text{fb}}(0)\big(y\big)$ in $\mathcal{D}(\mathcal{H}_P \otimes \mathcal{H}_C)$ and $\sigma (0) \in \mathcal{D}(\mathcal{H}_P \otimes \mathcal{H}_C)$, it holds that
\begin{equation}\label{tozero}
    \lim_{t \to +\infty}e^{\gamma\mathcal{L}_{\text{fb}}(0)\cdot t}E(0)=0.
\end{equation}
Because 
\begin{equation}
    \Vert \mathcal{U}_t e^{\gamma\mathcal{L}_{\text{fb}}(0)\cdot t}E(0)\Vert=\Vert e^{\gamma\mathcal{L}_{\text{fb}}(0)\cdot t}E(0)\Vert,
\end{equation}
it is clear that
\begin{equation}\label{thm3_14}
    \lim_{t \to +\infty} \Vert \mathcal{G}_{\gamma}(t,0)E(0)\Vert=0.
\end{equation}
At this point, the first term on the r.h.s of (\ref{thm3_8}) has been dealt with. We shall proceed to place an upper bound on the norm of the second term: $\int_0^t \mathcal{G}_{\gamma}(t,s) \mathcal{L}_{\text{noise}}\big(\sigma(s)\big)ds$. According to (\ref{thm3_12}), we have
\begin{multline}\label{thm3_15}
    \int_0^t \mathcal{G}_{\gamma}(t,s)\mathcal{L}_{\text{noise}}\big(\sigma (s)\big)ds\\
    =\int_0^t \mathcal{U}_t e^{\gamma\mathcal{L}_{\text{fb}}(0)\cdot (t-s)}\mathcal{U}_s^\dagger \mathcal{L}_{\text{noise}} \big(\sigma (s)\big)ds.
\end{multline}
For any superoperator $\mathcal{A}$ acting on $\mathcal{B}(\mathcal{H}_P\otimes \mathcal{H}_C)$, if $\mathcal{B}_0(\mathcal{H}_P\otimes \mathcal{H}_C)$ is $\mathcal{A}$-invariant, we denote the restriction of $\mathcal{A}$ on $\mathcal{B}_0(\mathcal{H}_P\otimes \mathcal{H}_C)$ by $\mathcal{A}|_0$. Note that, for $s \geq 0$, 
\begin{equation*}
    \text{tr}\Big(\mathcal{L}_{\text{noise}}\big(\sigma (s)\big)\Big)=0, 
\end{equation*}
which says that $\mathcal{L}_{\text{noise}}\big(\sigma (s)\big) \in \mathcal{B}_0(\mathcal{H}_P\otimes \mathcal{H}_C)$. Also, $\mathcal{U}_t$, $\mathcal{U}_s^\dagger$ and $e^{\gamma\mathcal{L}_{\text{fb}}(0)\cdot (t-s)}$ ($0\leq s \leq t$) are trace-preserving superoperators, which indicates that $\mathcal{B}_0(\mathcal{H}_P\otimes \mathcal{H}_C)$ is an invariant subspace under the operation of these superoperators. Therefore, we can write
\begin{multline}
    \mathcal{U}_t e^{\gamma\mathcal{L}_{\text{fb}}(0)\cdot (t-s)}\mathcal{U}_s^\dagger \mathcal{L}_{\text{noise}} \big(\sigma (s)\big)\\
    =\mathcal{U}_t |_0 e^{\gamma\mathcal{L}_{\text{fb}}(0)|_0\cdot (t-s)}\mathcal{U}_s^\dagger |_0 \mathcal{L}_{\text{noise}} \big(\sigma (s)\big),
\end{multline}
for $0\leq s\leq t$. It is then derived that
\begin{equation}\label{thm3_16}
    \begin{aligned}
        &\Vert \int_0^t \mathcal{U}_t e^{\gamma\mathcal{L}_{\text{fb}}(0)\cdot (t-s)}\mathcal{U}_s^\dagger \mathcal{L}_{\text{noise}} \big(\sigma (s)\big)ds \Vert \\
        \leq&\int_0^t\Vert\mathcal{U}_t |_0 e^{\gamma\mathcal{L}_{\text{fb}}(0)|_0\cdot (t-s)}\mathcal{U}_s^\dagger |_0 \mathcal{L}_{\text{noise}} \big(\sigma (s)\big)\Vert ds\\
        =&\int_0^t\Vert e^{\gamma\mathcal{L}_{\text{fb}}(0)|_0\cdot (t-s)}\mathcal{U}_s^\dagger |_0 \mathcal{L}_{\text{noise}} \big(\sigma (s)\big)\Vert ds\\
        \leq& \int_0^t\Vert e^{\gamma\mathcal{L}_{\text{fb}}(0)|_0\cdot (t-s)} \Vert \cdot \Vert\mathcal{U}_s^\dagger |_0 \mathcal{L}_{\text{noise}} \big(\sigma (s)\big)\Vert ds\\
        \leq &\int_0^t\Vert e^{\gamma\mathcal{L}_{\text{fb}}(0)|_0\cdot (t-s)} \Vert \cdot \Vert \mathcal{L}_{\text{noise}}\Vert ds.
    \end{aligned}
\end{equation}
It is shown in Appendix A that $\mathcal{L}_{\text{fb}}(0)|_0$ is a Hurwitz superoperator on $\mathcal{B}_0(\mathcal{H}_P\otimes \mathcal{H}_C)$. As a result, there exist $K>0$ and $\alpha >0$, 
\begin{equation}
    \Vert e^{\mathcal{L}_{\text{fb}}(0)|_0\cdot t}\Vert \leq K e^{-\alpha t},\quad t \geq 0.
\end{equation}
Consequently, 
\begin{equation}\label{thm3_17}
    \Vert e^{\gamma\mathcal{L}_{\text{fb}}(0)|_0\cdot t} \Vert \leq K e^{-\alpha \gamma t}, t \geq 0. 
\end{equation}
By (\ref{thm3_16}) and (\ref{thm3_17}), it is clear that
\begin{equation}\label{thm3_18}
    \begin{aligned}
        &\Vert \int_0^t \mathcal{G}_{\gamma}(t,s) \mathcal{L}_{\text{noise}} \big(\sigma (s)\big)ds \Vert \\
        \leq &\int_0^t\Vert e^{\gamma\mathcal{L}_{\text{fb}}(0)|_0\cdot (t-s)} \Vert \cdot \Vert \mathcal{L}_{\text{noise}}\Vert ds\\
        \leq &\int_0^t\Vert K e^{-\alpha \gamma (t-s)}\Vert \cdot \Vert \mathcal{L}_{\text{noise}}\Vert ds\\
        =&\frac{K}{\gamma \alpha} \Vert \mathcal{L}_{\text{noise}}\Vert (1-e^{-\alpha \gamma t}).
    \end{aligned}
\end{equation}
Then, based on (\ref{tozero}) and (\ref{thm3_18}), it is true that
\begin{multline}
    \limsup_{t \to +\infty} \Vert \mathcal{G}_{\gamma}(t,0)E(0)\Vert \\
    +\Vert \int_0^t \mathcal{G}_{\gamma}(t,s) \mathcal{L}_{\text{noise}} \big(\sigma (s)\big)ds \Vert \leq \frac{K}{\gamma \alpha} \Vert \mathcal{L}_{\text{noise}}\Vert.
\end{multline}
Furthermore, according to (\ref{thm3_8}),
\begin{equation}
    \limsup_{t \to +\infty} \Vert \sigma(t)-\lambda (t)\Vert \leq \frac{K}{\gamma \alpha} \Vert \mathcal{L}_{\text{noise}}\Vert.
\end{equation}
Since $\text{tr}_{\mathcal{H}_C}(\cdot)$ is a completely positive and trace-preserving operation, we have
\begin{equation}
    \Vert \text{tr}_{\mathcal{H}_C}\big(\sigma (t)\big)-\rho_D (t)\Vert \leq \Vert\sigma(t)-\lambda (t)\Vert,\quad t \geq 0.
\end{equation}
Therefore, in conclusion, $\limsup_{t \to +\infty} \Vert \text{tr}_{\mathcal{H}_C}\big(\sigma (0)\big)-\rho_D (t)\Vert \leq \frac{K}{\gamma \alpha} \Vert \mathcal{L}_{\text{noise}}\Vert$. $\hfill\square$
\end{pf}
{\bfseries Remark 3} \textit{Theorem 3 implies that the performance of noise suppression is dependent on the ratio of noise strength to feedback strength. Given a range of tolerance on performance, if the implemented coherent feedback protocol is sufficiently strong, then the performance will lie within the range, provided that noise strength is below a certain level. If the upper bound of noise strength is known, then it is possible to achieve better performance by increasing the feedback strength. In other words, the goal described by (\ref{hope}) can be achieved.}

{\bfseries Remark 4} \textit{In (\ref{plant_controller1}), the noise superoperator only involves coupling operators that act nontrivially only on the plant. However, the proof and conclusion of Theorem 3 are relevant with the norm of noise operator but not its nonlocality. In other words, even if} $\mathcal{L}_{\text{noise}}$ \textit{is associated with correlated coupling operators between the plant and controller, such noise can still be suppressed as indicated by (\ref{errorbound}).}

In fact, we believe that Theorem 3 also suggests the possibility of achieving noise suppression which is robust against inaccuracies in model realization. When implementing a coherent feedback protocol on the plant, the interaction Hamiltonian, the coupling operators on the controller, and even the plant Hamiltonian may not be perfectly realized as designed theoretically, which may result in an additional time-varying superoperator term in the system generator.

Suppose that the nominal system to be realized is described by (\ref{plant_controller11}). The actual system, however, may take the following form:
\begin{equation}\label{plant_controller12}
    \dot{\sigma}=\big(\mathcal{L}_{\text{p}} + \gamma \mathcal{L}_{\text{fb}}(t)+\mathcal{L}_{\text{noise}}+\mathcal{L}_{\text{unc}}(t)\big)(\sigma),
\end{equation}
where $\gamma >0$ and $\mathcal{L}_{\text{unc}}(t)$ absorbs the uncertainty in realizing $\mathcal{L}_{\text{p}}+\gamma \mathcal{L}_{\text{fb}}$. If there exists $L>0$, such that
\begin{equation*}
    \Vert \mathcal{L}_{\text{unc}}(t) \Vert \leq L,\quad t \geq 0,
\end{equation*}
for a very wide range of $\gamma$ and with $L$ not dependent on $\frac{\mathcal{L}_{\text{p}} + \gamma \mathcal{L}_{\text{fb}}(t)}{\Vert\mathcal{L}_{\text{p}} + \gamma \mathcal{L}_{\text{fb}}(t)\Vert}$, which implies that we are capable of engineering a wide range of systems with the same level of precision, then it is possible to achieve better performance by increasing feedback strength. In this case, by following a procedure similar to that in the proof of Theorem 3, it can be shown that there exists $K_1>0$ and $\alpha_1 >0$, such that
\begin{equation*}
    \limsup_{t \to +\infty} \Vert \text{tr}_{\mathcal{H}_C}\big(\sigma (t)\big)-\rho_D (t)\Vert \leq \frac{K_1}{\gamma \alpha_1} (\Vert \mathcal{L}_{\text{noise}}\Vert +L).
\end{equation*}

\subsection{A Possible Design}
Theorems 1, 2 and 3 have presented certain conditions under which our desired performance regarding noise suppression can be achieved. In this subsection, a possible coherent feedback design is presented which satisfies these conditions and thus fulfills our objective. An explicit construction of $\mathcal{L}_{\text{fb}}(t)$ is presented.

Let the dimensionality of the controller system be 2 ($\text{dim}(\mathcal{H}_C)=2$). An orthonormal basis of $\mathcal{H}_c$ is denoted by $\{|\nu_c^0\rangle,|\nu_c^1\rangle\}$. Only one coupling operator that acts nontrivially only on the controller is designed, which means that $N=1$ in (\ref{plant_controller3}). Let $\gamma$ be a positive number. The relevant operator $L_{C,1}$ is expressed as:
\begin{equation}\label{design1}
    L_{C,1}=\sqrt{\gamma}|\nu_c^0\rangle\langle\nu_c^1|.
\end{equation}
Moreover, let $\{|\nu_p^j\rangle\}_{j=0}^{N_p-1}$ be an orthonormal basis of $\mathcal{H}_P$, where $|\phi_0\rangle=|\nu_p^0\rangle$. Then, $H_I (0)$ is designed as:
\begin{multline}\label{design2}
     H_I (0)=\gamma\cdot\bigg(\sum_{k=0}^{\text{dim}(\mathcal{H}_P)-2}|\nu_p^k,\nu_c^1\rangle\langle\nu_p^{k+1},\nu_c^0|\\
     +\sum_{k=1}^{\text{dim}(\mathcal{H}_P)-1}|\nu_p^k,\nu_c^0\rangle\langle\nu_p^{k-1},\nu_c^1|\bigg),
\end{multline}
and
\begin{equation}\label{design3}
     H_I (t)=U_P (t) H_I (0) U_P^\dagger (t), \quad t \geq 0,
\end{equation}
with $U_P (t)$ given in (\ref{Up}). Here, $H_I (0)$ may be viewed as energy exchange between the plant and controller subsystems. We now give the following proposition.

{\bfseries Proposition 1} \textit{Suppose that the coherent feedback protocol $\mathcal{L}_{\emph{fb}}(\cdot)$ is designed following (\ref{design1}), (\ref{design2}) and (\ref{design3}). Then, if $\mathcal{L}_{\emph{noise}}=0$, there exists} $|\psi_0\rangle \in \mathcal{H}_C$ \textit{such that if} $\sigma(0)=|\phi_0\rangle\langle \phi_0|\otimes |\psi_0\rangle\langle \psi_0|$\textit{, eq. (\ref{thm1}) in Theorem 1 holds. Also, eq. (\ref{thm2}) in Theorem 2 is satisfied. In the case where} $\mathcal{L}_{\text{noise}}\neq 0$\textit{, eq. (\ref{errorbound}) in Theorem 3 holds.}

\begin{pf}
According to the statement of Theorems 1-3, it suffices to prove that the system $\dot{y}=\mathcal{L}_{\text{fb}}(0)\big(y\big)$ admits only one steady state in $\mathcal{D}(\mathcal{H}_P \otimes \mathcal{H}_C)$ with the form $|\phi_0\rangle\langle \phi_0|\otimes \rho_C$. 

Let us consider the following state 
\begin{equation}\label{prop1}
    \sigma_{\text{ini}}\triangleq |\phi_0\rangle\langle \phi_0|\otimes |\nu_c^0\rangle\langle\nu_c^0|.
\end{equation}
We shall prove that this is the unique steady state in $\mathcal{D}(\mathcal{H}_P \otimes \mathcal{H}_C)$.

It can be verified that the following equations hold:
\begin{equation}\label{prop2}
\begin{aligned}
    &[H_I (0),\sigma_{\text{ini}}]=0;\\
    &(I_P\otimes L_{C,1})\sigma_{\text{ini}}(I_P \otimes L_{C,1}^\dagger)=0;\\
    &(I_P\otimes L_{C,1}^\dagger L_{C,1})\sigma_{\text{ini}}=0;\\
    &\sigma_{\text{ini}}(I_P\otimes L_{C,1}^\dagger L_{C,1})=0.\\
\end{aligned}
\end{equation}
Therefore, it is true that $\mathcal{L}_{\text{fb}}(0)\big(\sigma_{\text{ini}}\big)=0$.

We then proceed to prove the uniqueness of the steady state $\sigma_{\text{ini}}$ in $\mathcal{D}(\mathcal{H}_P \otimes \mathcal{H}_C)$. Clearly, we have $\text{supp}(\sigma_{\text{ini}})=\text{span}\{|\phi_0,\nu_c^0\rangle\}$. According to (\ref{design1}) and (\ref{design2}), it also hold that
\begin{equation}
    (I_P\otimes L_{C,1}) |\phi_0,\nu_c^0\rangle=0,
\end{equation}
and
\begin{equation}
    \big(-\text{i}H_I (0)-\frac{1}{2 }I_P\otimes L_{C,1}^\dagger L_{C,1}\big) |\phi_0,\nu_c^0\rangle=0.
\end{equation}
Therefore, $\text{span}\{|\phi_0,\nu_c^0\rangle\}$ is a common invariant subspace of both $I_P\otimes L_{C,1}$ and $-\text{i}H_I (0)-\frac{1}{2}I_P\otimes L_{C,1}^\dagger L_{C,1}$. We shall prove next, by contradiction, that there exists no nonzero common invariant subspace of $I_P\otimes L_{C,1}$ and $-\text{i}H_I (0)-\frac{1}{2}I_P\otimes L_{C,1}^\dagger L_{C,1}$ in $\text{span}\{|\phi_0,\nu_c^0\rangle\}^\perp$.

Suppose there is a nonzero common invariant subspace of $I_P\otimes L_{C,1}$ and $-\text{i}H_I (0)-\frac{1}{2}I_P\otimes L_{C,1}^\dagger L_{C,1}$ in $\text{span}\{|\phi_0,\nu_c^0\rangle\}^\perp$, which is denoted by $V_1$. Then, $\forall x \in V_1$, it must be the case that 
\begin{equation}
    |\phi_0,\nu_c^1\rangle\langle\phi_0,\nu_c^1|x=0.
\end{equation}
Otherwise, it can be checked that
\begin{multline}
    |\phi_0,\nu_c^1\rangle\langle\phi_0,\nu_c^1|(I_P\otimes L_{C,1})x\\
    =|\phi_0,\nu_c^1\rangle\langle\phi_0,\nu_c^1|x\neq 0,
\end{multline}
which says that $(I_P\otimes L_{C,1})x$ is not in $\text{span}\{|\phi_0,\nu_c^0\rangle\}^\perp$, and thus not in $V_1$, contradicting the supposition that $V_1$ is $(I_P\otimes L_{C,1})$-invariant. Therefore, it has been shown that $V_1 \perp \text{span}\{|\phi_0,\nu_c^1\rangle\}$.

Next, $\forall x \in V_1$, it must be true that
\begin{equation}
    |\nu_p^1,\nu_c^0\rangle\langle\nu_p^1,\nu_c^0|x=0.
\end{equation}
Otherwise, it is derived that
\begin{multline}
    |\phi_0,\nu_c^1\rangle\langle \phi_0,\nu_c^1|\big(-\text{i}H_I (0)-\frac{1}{2}I_P\otimes L_{C,1}^\dagger L_{C,1}\big) x\\
    =-\text{i}|\nu_p^1,\nu_c^0\rangle\langle\nu_p^1,\nu_c^0|x\neq 0,
\end{multline}
which indicates that $\big(-\text{i}H_I (0)-\frac{1}{2}I_P\otimes L_{C,1}^\dagger L_{C,1}\big)x$ is not in $\text{span}\{|\phi_0,\nu_c^0\rangle, |\phi_0,\nu_c^1\rangle\}^\perp$, and thus not in $V_1$, contradicting the supposition that $V_1$ is $\big(-\text{i}H_I (0)-\frac{1}{2}I_P\otimes L_{C,1}^\dagger L_{C,1}\big)$-invariant. Therefore, it has been further shown that $V_1 \perp \text{span}\{|\nu_p^1,\nu_c^0\rangle\}$.

By following similar procedures and combining what we have already proved, it is can be shown that $\forall 1\leq i\leq \text{dim}(\mathcal{H}_P)-1$ and $j=0,1$, $V_1 \perp \text{span}\{|\nu_p^i,\nu_c^j\rangle\}$, and also $V_1 \perp \text{span}\{|\nu_p^1,\nu_c^0\rangle\}$. Since $V_1 \perp \text{span}\{|\nu_p^0,\nu_c^0\rangle\}$, it must hold that $V_1=\{0\}$, which contradicts the supposition that $V_1$ is nonzero. 

Based on what has been shown so far and according to \cite{TICOZZI20092002}, it is true that, for any $\rho_0 \in \mathcal{D}(\mathcal{H}_P\otimes \mathcal{H}_C)$, 
\begin{equation}
    \lim_{t\to +\infty} \text{tr}\big(|\phi_0,\nu_c^0\rangle\langle \phi_0,\nu_c^0|\}e^{\mathcal{L}_{\text{fb}(0)}\cdot t}\rho_0\big)=0,
\end{equation}
which implies that there are no other steady states than $|\phi_0,\nu_c^0\rangle\langle \phi_0,\nu_c^0|$ in $\mathcal{D}(\mathcal{H}_P\otimes \mathcal{H}_C)$. The proof is completed. $\hfill\square$
\end{pf}

\section{Example}
In this section, an example is presented to illustrate our results. Let us consider a plant of two qubits. Therefore, we set $\mathcal{H}_P=\mathbb{C}^2\otimes \mathbb{C}^2$. Let $\{|0\rangle,|1\rangle\}$ be an orthonormal basis of $\mathbb{C}^2$, where
\begin{equation}
    |0\rangle\triangleq \begin{pmatrix}0\\1\end{pmatrix},\quad |1\rangle\triangleq \begin{pmatrix}1\\0\end{pmatrix}.
\end{equation}
An orthonormal basis of $\mathcal{H}_P$ can thus be expressed as
\begin{equation}
    \{|00\rangle , |01\rangle , |10\rangle , |11\rangle \}.
\end{equation}
The initial state in (\ref{plant}) is set as $|00\rangle$, and the Hamiltonian $H_P$ in (\ref{plant}) is chosen as $\sigma_x \otimes \sigma_x$, where
\begin{equation}
    \sigma_x \triangleq \begin{pmatrix}0&1\\1&0\end{pmatrix}
\end{equation}
denotes the Pauli-X matrix. This choice of the initial state and Hamiltonian indicates that our desired trajectory satisfies: 
\begin{equation}\label{eg1}
\dot{\rho_D}=-\text{i}[\sigma_x \otimes \sigma_x,\rho_D],\quad \rho_D (0)=|00\rangle\langle 00|,
\end{equation}
and
\begin{equation}\label{eg2}
    \rho_D (t)=e^{-\text{i}(\sigma_x \otimes \sigma_x) t}|00\rangle\langle 00| e^{\text{i}(\sigma_x \otimes \sigma_x) t}.
\end{equation}

If the ``genuine stochasticity" hypothesis regarding quantum measurements is true in reality, then any quantum system may act as a Quantum Random Number Generator (QRNG). The plant considered in this section may act as a QRNG with four outcomes.  The distribution of outcomes depends on the time at which measurement is performed. Suppose that we would like to use this plant as a QRNG. Then, if measurement is performed at $t_1>0$ with measurement basis $\{|00\rangle,|01\rangle,|10\rangle,|11\rangle\}$, we expect the probabilities of the four outcomes to be $\langle 00|\rho_D (t_1)|00\rangle$, $\langle 01|\rho_D (t_1)|01\rangle$, $\langle 10|\rho_D (t_1)|10\rangle$ and $\langle 11|\rho_D (t_1)|11\rangle$, respectively. By choosing different measurement times, different distributions may be generated. 

However, if the plant is affected by noise (transient or persistent), the probabilities of outcomes may deviate from expected values. By suppressing noise, we may be able to reduce the error regarding outcome probabilities, therefore making the QRNG more reliable. According to the results of this article, coherent feedback may be helpful.

The controller is set to be a two-level system, which says that $\mathcal{H}_C =\mathbb{C}^2$. The coherent feedback protocol is designed following (\ref{design1}), (\ref{design2}) and (\ref{design3}), so that the conditions in Theorems 1, 2 and 3 are satisfied according to Proposition 1. That is, for $\gamma >0$, we design:
\begin{equation}
    L_{C,1}=\sqrt{\gamma}|0\rangle\langle 1|,
\end{equation}
and
\begin{equation}
    H_I (0)=\gamma\!\begin{pmatrix}0&0&0&0\\1&0&0&0\\0&1&0&0\\0&0&1&0\end{pmatrix}\! \otimes \! |1\rangle\langle 0|+\gamma \!\begin{pmatrix}0&1&0&0\\0&0&1&0\\0&0&0&1\\0&0&0&0\end{pmatrix}\! \otimes \! |0\rangle\langle 1|,
\end{equation}
and 
\begin{equation}
    H_I (t)=e^{-\text{i}(\sigma_x \otimes \sigma_x \otimes I_2) t}H_I (0) e^{\text{i}(\sigma_x \otimes \sigma_x \otimes I_2) t},\quad t \geq 0.
\end{equation}

The case where $\mathcal{L}_{\text{noise}}=0$ is considered first. Our simulation is related to the following scenario. At $t=0$, the plant-controller composite system is initialized in $|000\rangle\langle 000|$ and with coherent feedback protocol prescribed in this section, the system begins to evolve. However, at $t=t_a=1$, the state of the composite system is suddenly affected by a decoherence noise action $\mathcal{T}_{\text{decohere}}$, such that
\begin{equation}
    \mathcal{T}_{\text{decohere}}(\cdot)\triangleq \sum_{i,j,k=0,1}|i,j,k\rangle\langle i,j,k|(\cdot)|i,j,k\rangle\langle i,j,k|.
\end{equation}
After changing the system state, the noise disappears at $t_a=1$. Next, we denote
\begin{equation}
    D(t)\triangleq \frac{1}{2} \Vert \text{tr}_{\mathcal{H}_C}\big(\sigma (t)\big)-\rho_D (t) \Vert,\quad t \geq 0,
\end{equation}
which is the trace distance between the plant's real-time state (which is reduced from the composite system state) and the state at time $t$ on the desired trajectory. The variation of $D(t)$ against $t$ is simulated ($\gamma=5$), and the associated result is shown in Fig.1.
\begin{figure}
    \centering
    \includegraphics[height=6.5cm]{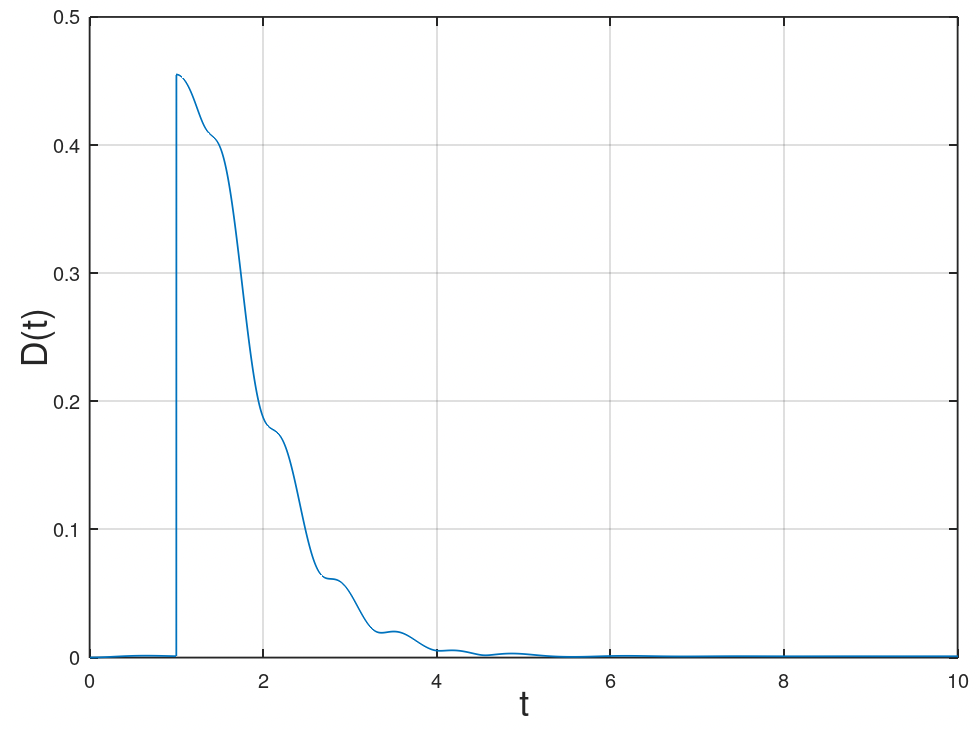}    
\caption{Simulated variation of distance between plant's state and desired state.}  
 \label{fig1}
\end{figure}

In Fig.1, it is seen that, for $0\leq t <t_a =1$, the curve nearly completely overlaps the $t$-axis. This observation is viewed together with Theorem 1. Note that, in our setting here, initialization is perfect and during this time interval there are no noise acting on the system. Theorem 1 says that the plant state evolution should exactly follow the desired trajectory. 

Moreover, after the noise action, it is seen that the curve shows a general tendency to approach the $t$-axis. This observation is viewed in conjunction with Theorem 2, which says that $D(t)$ should tend to zero in this case. 

Then, we proceed to the case where $\mathcal{L}_{\text{noise}}\neq 0$. The following three coupling operators are considered \big($M=3$ in (\ref{plant_controller4})\big):
\begin{equation}\label{eg3}
    \begin{aligned}
        &L_{P,1}=0.5|0\rangle\langle 1| \otimes I_2;\\
        &L_{P,2}=0.5 I_2 \otimes \sigma_z;\\
        &L_{P,3}=0.5\sigma_x \otimes |0\rangle\langle 1|,
    \end{aligned}
\end{equation}
where
\begin{equation}
    I_2\triangleq \begin{pmatrix}1&0\\0&1\end{pmatrix},\quad \sigma_z \triangleq \begin{pmatrix}1&0\\0&-1\end{pmatrix}.
\end{equation}
Moreover, initialization is not considered to be perfect in this case. The initial state chosen as:
\begin{multline}\label{eg4}
     \sigma (0)=(0.8|00\rangle\langle 00|+0.1 |01\rangle\langle 01|+0.05|10\rangle\langle 10|\\+0.05|11\rangle\langle 11|)\otimes(0.9|0\rangle\langle 0|+0.1|11\rangle\langle 11|).
\end{multline}

\begin{figure}
    \centering
    \includegraphics[height=6.5cm]{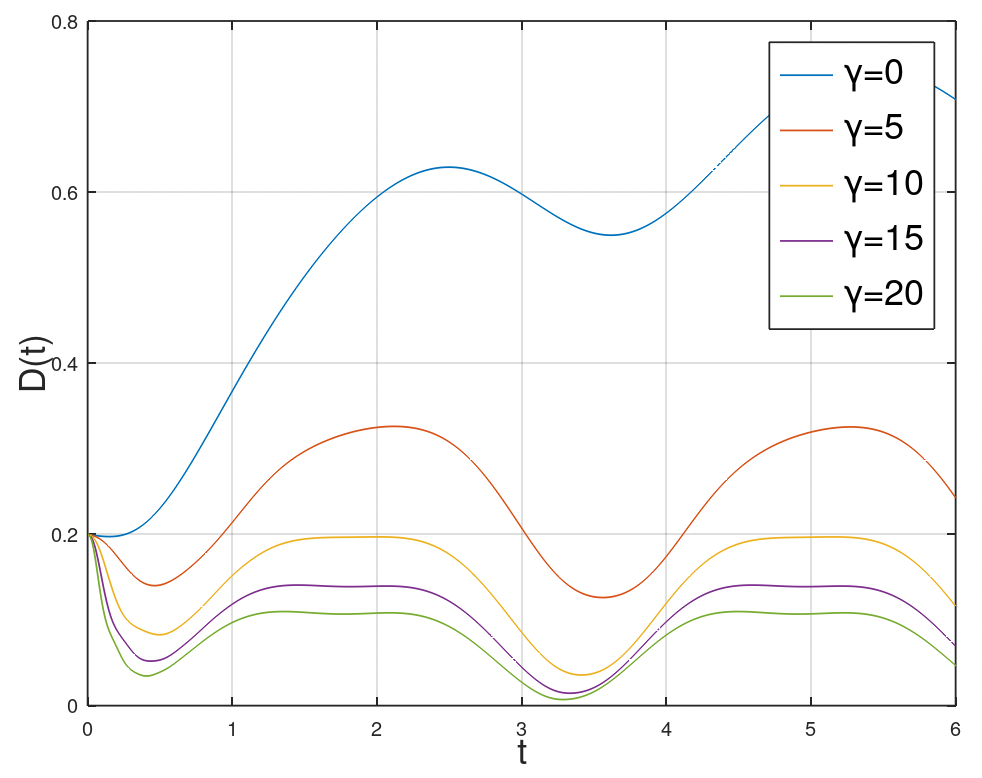}    
\caption{Simulated variation of distance between plant's state and desired state, w.r.t five choices of $\gamma$.}  
 \label{fig2}
\end{figure}
   
With $\mathcal{L}_{\text{fb}}$ determined by the coupling operators in (\ref{eg3}) and initial state set as (\ref{eg4}), we simulate the variation of $D(t)$ against $t$ for 5 choices of $\gamma$, namely, $\gamma=0,5,10,15,20$. Clearly, the $\gamma=0$ scenario represents the absence of feedback, and as $\gamma$ increases, the feedback strength increases. The simulation result is shown in Fig.2.

From Fig.2, we observe that the feedback-absent curve deviates from the $t$-axis substantially, which makes sense given the presence of persistent noise. However, as feedback strength increases, it looks as if there were a force pressing the curve towards the $t$-axis. This observation is viewed with Theorem 3, which implies improved noise suppression performance as feedback grows stronger.

\section{Conclusion}
We have described the problem of achieving noise suppression with coherent quantum feedback. Several conditions on the coherent feedback protocol that ensures desired noise suppression performance are presented, and an explicit protocol design which satisfies these conditions are given. An example regarding a two-qubit plant and a two-level quantum controller is also given.

\begin{ack}                          
This work is partially financially supported by Innovation Program for Quantum Science and Technology 2023ZD0300600, Guangdong Provincial Quantum Science Strategic Initiative (No. GDZX2200001), Hong Kong Research Grant Council (RGC) under Grant No. 15213924, National Natural Science Foundation of China under Grants No. 62173288.
\end{ack}


\bibliographystyle{apalike}        
\bibliography{autosam}           



\appendix
\section{Proof that $\mathcal{L}_{\text{fb}}(0)|_0$ is Hurwitz}    
Note that this proof relies on the conditions in Theorem 3. Let us denote the subspace of $\mathcal{B}(\mathcal{H}_P \otimes \mathcal{H}_C)$ spanned by the nonzero eigenoperators of $\mathcal{L}_{\text{fb}}(0)$ and their associated generalized eigenoperators as $V_{\text{n}0}$. We shall prove that $V_{\text{n}0}=\mathcal{B}_0(\mathcal{H}_P \otimes \mathcal{H}_C)$.

Since the dynamical system $\dot{y}=\mathcal{L}_{\text{fb}}(0)\big(y\big)$ admits a unique steady state in $\mathcal{D}(\mathcal{H}_P \otimes \mathcal{H}_C)$, the kernel of $\mathcal{L}_{\text{fb}}(0)$ is one-dimensional and 
\begin{equation}
    \text{dim}(V_{\text{n}0})=\big(\text{dim}(\mathcal{H}_P) \cdot \text{dim}(\mathcal{H}_C) \big)^2-1.
\end{equation}
Also, for any nonzero element $X$ of $V_{\text{n}0}$, it holds that $\text{tr}(X)=0$. Since the dimension of $\mathcal{B}_0(\mathcal{H}_P \otimes \mathcal{H}_C)$ is $\big(\text{dim}(\mathcal{H}_P) \cdot \text{dim}(\mathcal{H}_C)\big)^2-1$, the basis of $V_{\text{n}0}$ is also a basis of $\mathcal{B}_0(\mathcal{H}_P \otimes \mathcal{H}_C)$. Therefore, $V_{\text{n}0}=\mathcal{B}_0(\mathcal{H}_P \otimes \mathcal{H}_C)$.

Moreover, $\mathcal{L}_{\text{fb}}(0)$ has no eigenvalues with positive real parts and no purely imaginary eigenvalues. The same must hold for its restriction on $\mathcal{B}_0(\mathcal{H}_P \otimes \mathcal{H}_C)$, which is $\mathcal{L}_{\text{fb}}(0)|_0$. Next, $\mathcal{L}_{\text{fb}}(0)|_0$ must not have a zero eigenvalue. If it had one, then the corresponding eigeneoperators could not be linearly expressed by the elements of $\mathcal{B}_0(\mathcal{H}_P \otimes \mathcal{H}_C)$. In conclusion, $\mathcal{L}_{\text{fb}}(0)|_0$ is Hurwitz.

\section{Discussion on Realization of Time-Dependent Hamiltonian}
In this article, Theorems 1-3 include (\ref{TDH}) as a part of their statements, which indicates that the realization of time-dependent Hamiltonian (\ref{TDH}) is an important issue. Here, we present a discussion on its realization, which may lead to a topic for future research.

Let us focus on the case where the plant and controller are all collections of qubits. For a collection of $N$ qubits, the underlying Hilbert space is $\mathcal{H}=\mathbb{C}^{2^N}$. Any time-dependent Hamiltonian $H(t)$ on $\mathcal{H}$ can be decomposed as:
\begin{equation}
H(t)=\sum_{j_1,...,j_N}c_{j_1,...,j_N}(t)\bigotimes_{i=1}^N\sigma_{j_i},
\end{equation}
where $j_i \in \{0,x,y,z\}$ ($\sigma_0 \triangleq I_2$). One may expect significant challenges in realizing $H(t)$, especially if $N$ is large, since engineering many-qubit interactions may be difficult, and the physical meaning of their linear combinations may be unclear. Nevertheless, endeavoring to experimentally achieve this, perhaps in the future, still might be worthwhile and rewarding, since we will have more flexibility in quantum engineering design. 

However, approximating the unitary contribution that $H(t)$ generates may also be acceptable. In classical digital control, a continuous control action may be discretized. Similarly, we may divide the time interval on which a control protocol is to act into fine grids and find approximations of the actual unitary contribution on each grid. Then, each unitary contribution may be decomposed into elementary quantum gates. Since single-qubit and two-qubit gates already form a universal gate set, we may not need to engineer many-qubit interactions in this case. We expect that (not making a rigorous claim though), as the number of grids tends to infinity, the approximated dynamics may converge to the real dynamics. 

In \cite{PhysRevA.91.012118,ROUCHON2022252}, discretized formulations have been considered. It remains to be seen if such formulations, and possibly other methods, may be suitable for modeling the dynamics with discrete quantum gates replacing a continuous time-dependent Hamiltonian. Moreover, relevant convergence analysis may be an interesting topic for future research.

\end{document}